# A Bio-Polymer Transistor: Electrical Amplification by Microtubules

Avner Priel[1†], Arnolt J. Ramos[2], Jack A. Tuszynski[1], and Horacio F. Cantiello[2]

[1]*Department of Physics, University of Alberta, Edmonton, AB T6G 2J1, Canada.* [2]*Renal Unit, Massachusetts General Hospital, and Harvard Medical School, Charlestown, Massachusetts, USA,*

[†]Corresponding author. Avner Priel, Department of Physics, University of Alberta, Edmonton, AB T6G 2J1, Canada. Tel: (780) 492-3579. Email: apriel@phys.ualberta.ca
Horacio F. Cantiello, Maassachusetts General Hospital, Charlestown, Massachusetts, USA, Email: cantiello@helix.mgh.harvard.edu

## Abstract

Microtubules (MTs) are important cytoskeletal structures, engaged in a number of specific cellular activities, including vesicular traffic, cell cyto-architecture and motility, cell division, and information processing within neuronal processes. MTs have also been implicated in higher neuronal functions, including memory, and the emergence of "consciousness". How MTs handle and process electrical information, however, is heretofore unknown. Here we show new electrodynamic properties of MTs. Isolated, taxol-stabilized microtubules behave as bio-molecular transistors capable of amplifying electrical information. Electrical amplification by MTs can lead to the enhancement of dynamic information, and processivity in neurons can be conceptualized as an "ionic-based" transistor, which may impact among other known functions, neuronal computational capabilities.

## Introduction

MTs are long hollow cylinders made up of GTP-dependent αβ-tubulin dimer assemblies (1). MTs have outer diameters of approximately 25 nm and inner diameters of 15 nm, with lengths that reach several micrometers. Recently, it has become apparent that neurons may utilize MTs in cognitive processing. MT-associated proteins, including tau and MAP2, have been implicated in such neuronal processes as learning and memory (2-4). MTs are also linked to the regulation of a number of ion channels, thus contributing to the electrical activity of excitable cells (5). Thus, MTs may play a heretofore, unknown role in the processing of electrical signals within the neuron. Morphologically speaking, the elongating processes that sprout from the neuronal soma include a single axon and a variable number of dendrites, both take cable-like properties and contain abundant MTs. Dendrites, in particular, are responsible for both synaptic integration and synaptic plasticity, essential functions of neuronal activity (6,7). It is increasingly clear, that cytoskeletal structures play a central role in both the process of ramification, and more specialized neuronal functions, including neuronal processivity, long-term potentiation (LTP), long-term depression (LTD), and memory formation (8,9). Cytoskeletal alterations may reflect changes of neural circuits in response to learning and experience, and they must involve highly dynamic regulatory mechanisms. How this cytoskeletal control of neuronal plasticity takes form, remains ill defined. Interestingly, a century ago Cajal proposed that the "cytoskeleton contain a system for the conduction of the nerve impulse" (10). To elucidate the electrodynamic properties of MTs and gain insight into the regulatory role that MTs play in dendritic activity, we assessed whether MTs are capable of processing electrical information.

## Methods

### Preparation of isolated microtubules

Tubulin was polymerized with a combined GTP-taxol "hybrid" protocol, following loosely general guidelines as described in Mitchinson Lab's online protocols (http://mitchison.med.harvard.edu/protocols/). Briefly, all reactions were conducted in BRB80 solution, containing 80 mM PIPES, 1 mM $MgCl_2$, 1 mM EGTA, and pH 6.8 with KOH. In some experiments 10 μl glycerol was added prior to mixing the tubulin. An aliquot of tubulin in solution (Catalog #T238, Cytoskeleton, Denver, CO) was mixed with 1 mM DTT and 1 mM GTP in an ice bed, and incubated for five min. Taxol (paclitaxel, Sigma) was then added in steps, first from a 1 μM, and then 10 μM solutions. Each step-wise addition of taxol was followed by incubation for 20-30 min in a water bath at 37 °C. Taxol was kept in a 10 mM stock in anhydrous DMSO. In some instances, the tubulin-containing solution was mixed with BRB80 solution (x2) and added 2 mM DTT, 2 mM GTP and 20% DMSO, followed by incubation at 37 °C for 20-30 min. Polymerized tubulin was centrifuged in a microfuge for 30 min at 14,000 rpm. Two-thirds of the supernatant was discarded, and the remainder constituted the microtubule sample, which was further stabilized by addition of 1 mM GTP and 1 μM taxol. Microtubules were visualized under phase contrast microscopy (IMT-2, Olympus, x60) to ascertain the lack of nucleated centers induced by higher concentrations of taxol. Microtubules were stable at room temperature for up to two days.

## Measurement of electrical signals in isolated microtubules

Electrical data were collected with a modified dual "patch-clamp" setup as previously reported (14). Briefly, the electrical setup consisted of two independent patch-clamp amplifiers, PC501 and PC501A (Warner Instruments, Hamden, CT), with 1 G$\Omega$ and 10 G$\Omega$ feedback resistors, respectively, one to stimulate the isolated microtubule (stimulus, s) and the other to collect current signals (collection, c). Electrical signals were simultaneously collected by respective analog-inputs of an A/D converter (TL-1 DMA Interface, Axon Instruments, Inc., Foster City, CA). Voltage stimulus protocols were constructed in Clampex 5.5.1 (Axon Instruments) ran from a computer. The sampling intervals varied depending on the protocol, such that in most cases a 100-msec signal length contained at least 400 samples. Offset tip potentials; within the range of a few mV were compensated immediately prior to data acquisition. Further zeroing of the offset signal was conducted off-line during data processing. Signals were Gaussian-filtered at a cutoff frequency of 2.2 kHz. Pipettes were made from borosilicate capillaries (1.2 mm external diameter), and were pulled with tip diameters below 3 $\mu$m. The pipettes contained, and were immersed in a solution containing in mM: 135 KCl, 5 NaCl, 0.8 MgCl$_2$, 1.2 CaCl$_2$, 10 Hepes. Occasionally, the pipette tip contained a 1:10 dilution of the taxol solution to help attach the microtubule. The pipettes were connected to the electrodes and head-stage of their respective patch-clamp amplifiers. Agar bridges containing 3% agar in 150 mM KCl solution grounded both amplifiers to the chamber. The circuit was closed by two AgCl plated Ag ground electrodes in solutions far away (1-2 cm) from the pipette tips.

## Data analysis

Data were acquired at 2 KHz and further filtered off-line for display and analysis when required. Noise analysis was conducted as Lorentzian spectral analysis with a subroutine of Axograph 4.0 (Axon Instruments) from unfiltered files where the variance vs. frequency was plotted for paired experiments before and after MT-attachment was obtained in the experiment. The method for deriving the amplification is different for the square pulse and for the triangle wave. In the case of square pulse, we evaluate the average amplitude in solution only and with MT attached for the same conditions. The amplification is simply the ratio. In the case of triangle pulses a different strategy was used, using the slope of the linear regression of the MT-attached data as a function of the solution-only data at the pulse times. Data are expressed as n = number of files averaged ± SEM, where "t" test was used to assess statistical significance at p<0.05.

## Results

Microtubules were formed and stabilized as indicated in Methods. Isolated MTs were visualized and identified under phase contrast microscopy, and connected to the tips of two patch clamp amplifiers (Fig. 1a), such that electrical stimulation could be applied to one of them, while the elicited signal could be collected with the other one (Fig. 1b). MTs were electrically stimulated by applying 5-10 msec input voltage pulses with amplitudes in the range of ±200 mV. The resulting electrical signals were obtained at the opposite end of an MT, 20-50 μm away, with a pipette also connected to a patch amplifier, which was kept "floating" at zero mV. We observed two remarkable findings. First, coupling of the stimulus pipette to a MT, increased the overall conductance of the pipette by more than 300%. The stimulus pipette resistance in KCl saline (135 mM) was 21.6 ± 0.6 MΩ (n=16). The coupling ratio between pipettes in solution was 41 ± 12% before attachment to an MT. Thus, the collection pipette "read" ~40% of the stimulus in solution. The evoked current increased from 1.91 ± 0.13 nA (n=16) to 2.78 ± 0.17 nA (n=35) after attachment (Fig. 2a). The signals reaching the collection site were in 30/35 of the cases higher than those obtained in free solution, indicating that the MT amplified both the electrical pulse injected at the stimulus site and the collection site as well (Fig. 2a). Transfer amplification ratios were up to 2.35 with an average of 1.69 ± 0.06 (n=30, Fig. 2c). Thus, MTs improve electrical connectivity between two locations in saline solution. The currents measured at the collection site were linearly dependent on the stimulus pipette input voltage, indicating a strictly inverse ohmic response, i.e. linear amplification (Fig. 2b). The MT conductances reached up to 9 nS, much higher than that expected from channel conductances (5-200 pS, Fig. 2b, Inset). The electrical amplifying effect of the MT was observed in either direction (stimulus by either amplifier). To further prove the linear response in amplification as well as the speed of the electrical amplifying phenomenon, voltage ramps were also applied within the range of ±100 mV, at two different rates (100 V/sec and 20 V/sec, Fig. 2b). The stimulated currents at either rate were again linear, showing a remarkable reproducibility within the range of voltages studied. The amplification ratio remained constant within the range of voltages and rates tested, and the noise in the signal was, as expected increased as well. The fastest ramps tested showed, however, a 40 μsec (or shorter) delay between the stimulus and collection sites, which was consistent with the shift, in voltage decay after pulse stimulation (Fig. 2a, Right). This suggests a lower bound for the transfer rate of the electrical signal of the order of 1.0 m/sec. In two out of seven experiments, electrical amplification further increased by addition of GTP (1 mM) to the bath solution (data not shown).

## Discussion

Our findings demonstrate that electrical amplification by MTs is equivalent to the polymer's ability to act as a bio-molecular transistor at its core. Based on a minimalistic model, both a constant electrical polarization, and an active component are required. Despite current interest and extensive literature on the properties and functions of MTs, practically nothing is known about the ionic properties of the MTs intra-polymeric compartment. However, molecular dynamics simulation of tubulin structure (11,12) indicates a strong negative surface charge distribution of the order of 20e$^-$ per monomer, distributed more on the outer surface than in the inner core with ratio of ~2:1. Our findings support these theoretical results manifested in the model via a constant (permanent) electric polarization, which follows localized Nernst potentials arising from asymmetries in the ionic distributions between the intra- and extra-MT environments. This polarization is modulated by electrical stimulation such that the forward-reverse biased junctions of an intra-molecular transistor (Fig. 3), creates a proper MT-adjacent ionic cloud environment, which allows amplification of axially transferred signals. The proposed model implies that intrinsic semi-conductive like properties of the structured tubulin dimers, are such that an effective transistor is being formed (13), whose gating ability to modulate localized charges may help amplify axial ionic movements. Recent studies aimed at solving the PB equation for a section of an MT in ionic conditions similar to those reported here (i.e. 150 mM saline) (11), suggest a highly ordered charge distribution. Positive charges are "sandwiched" within a two-layered structure of negative charges facing either side of the MT. There also is an asymmetrical distribution of charges on the plus-minus orientation of the MT. However, to better understand the dynamical properties of the electric field around the MT it is imperative to apply a method that incorporates the diffusive part of the system. A possible candidate is the Poisson-Nernst-Planck (PNP) theory in which both the electrical potential generated by fixed charges and the flux generated by the moving charges due to potential and concentration gradients, are taken into account. This method has been applied successfully to describe ion channels (29) and it would be interesting to test the applicability of this method to a system as large as the microtubule.

An electrical amplifier by a cytoskeletal biopolymer such as the MT provides a relevant and important key component to an electrodynamic network in conjunction with electrical coupling by current generators (i.e., MT-regulated channels (5)), and MT-coupled intracellular transmission lines (F-actin) (14,15). It is therefore plausible that electrical amplification by MTs may play an important role in processing of electrical information in neuronal function. Our data further indicate that this phenomenon is accomplished with speed at least as high as those reported for cable properties in neurites (16).

MT networks (MTN) play relevant roles in neuron formation and function (17-19). Interconnected bundles of MTs for example, are present in the axon, the axon hillock, and the dendritic shaft, all regions where ion channels are found. In particular, dendritic MTs are arranged in grid-like networks of mixed polarity interconnected by MAP2s. Thus, we envision a mechanism in which, MT-ion channel interactions may regulate synaptic plasticity by the MT's ability to allow spatially oriented intracellular electrical signals in connection with actin filaments, also able to transmit electrical information (14,15). According to this hypothesis postsynaptic electrical signals elicit ion waves along the associated actin filaments at the synaptic spine that propagate to the MTN where they serve as input signals. The MTN, operating as a large high dimensional "state machine", evolves these input states, e.g., by supporting non-linear wave collisions. The output from the MTN is the state of the system that is being "read" (sensed) to propagate and electrically stimulate remote voltage-sensitive ion-channels. Thus, our findings provide several advantages in the context of neural function. Cable theory analysis of dendrites (20,21) has challenged the simple integrate and fire models suggesting that dendrites impose a heavy conductance load on the soma, acting as low-pass filters of post synaptic potentials, hence changing the response to synaptic activities (21). The dendritic electrical properties are dynamically changed by modulation of voltage-gated ion channels (6,22,23), and by changes in cytoskeletal structures (24-26). Thus, MT electrical amplification may be central to revised models of neuronal adaptability (22,27,28) providing renewed support to nonlinear models of neuronal activity.

**Acknowledgements**

HC wishes to acknowledge Itsushi Minoura and Etsuko Muto, RIKEN Brain Science Institute, Japan, with whom preliminary studies were conducted. Funding from NSERC (Canada), MITACS and Technology Innovations, LLC of Rochester, NY supported this research. We thank one of the reviewers for bringing the relevance of the PNP theory to our attention.

# Figure Legends

**Figure 1.**

Electrical setup. **a. (Top)** A free-floating MT was "connected" to two patch pipettes. Horizontal bar is 50 μm. (**Bottom**) Electrical signals were applied to one end, and collected from the other end with two patch clamp amplifiers. **b.** voltage pulse (+100 mV) was first applied to the "stimulus" pipette in saline solution. The recorded current from stimulus pipette (Left-Bottom) and electrical coupling by the "collection" pipette (Left-Top) were obtained. A similar approach was repeated after MT attachment. Currents were higher after attachment to an MT (Right), compared to saline solution (Left). Representative tracings from 16 experiments are shown.

**Figure 2.**

Pulse stimulation of isolated microtubules. **a.** Average electric pulses (100 mV, 5 msec) for the "stimulus" (Red) and the "collection" (Blue) sites. Data on the Left indicate the signal before attachment. Values are the mean ± SEM for five pipettes with similar electrical properties. The decay response of the MT-connected tip at the end of the electrical pulse was ~40 μsec faster after connection (Right). **b.** MT-attached signals on the collection site for 5 msec ramps at different voltages. Fitted linear slopes (Green) indicate a linear response. Inset. Current-to-voltage relationship for a connected MT (open circles) compared to free solution (filled circles). The conductance is linear. The signals increased in average by 1.69, at an average distance of 35 μm. **c.** Correlation of amplification ratios between the stimulus and collection sites. Identity line (Red) indicates no amplification by the connected MT. Amplification factors are shown for either 10 msec (f1) and 5 msec (f5) ramps, for positive (P) and negative (M) potentials (last three digits).

**Figure 3.**

Electrical model of the MT. **a.** Effective electrical model of the MT (Right) is consistent with an electrical amplifier, including a molecular "bio-transistor", and energy sources in the form of batteries to polarizing the amplifier. **b.** Relation of electrical model to the electrostatic properties of an MT, for both the plus and minus ends, potential isocontours, reproduced from Baker et al. (11). Electrostatic profile of a cross-section of the MT (**Right**), shows periodic distribution of positive charges (Black) on the surface of the electronegative MT. Periodically distributed charges (large and small pentagons), maintain a "band-junction" in the surrounding counterions. This is based on an electric potential difference in the MT wall (open circles and filled circles for cations and anions, respectively).

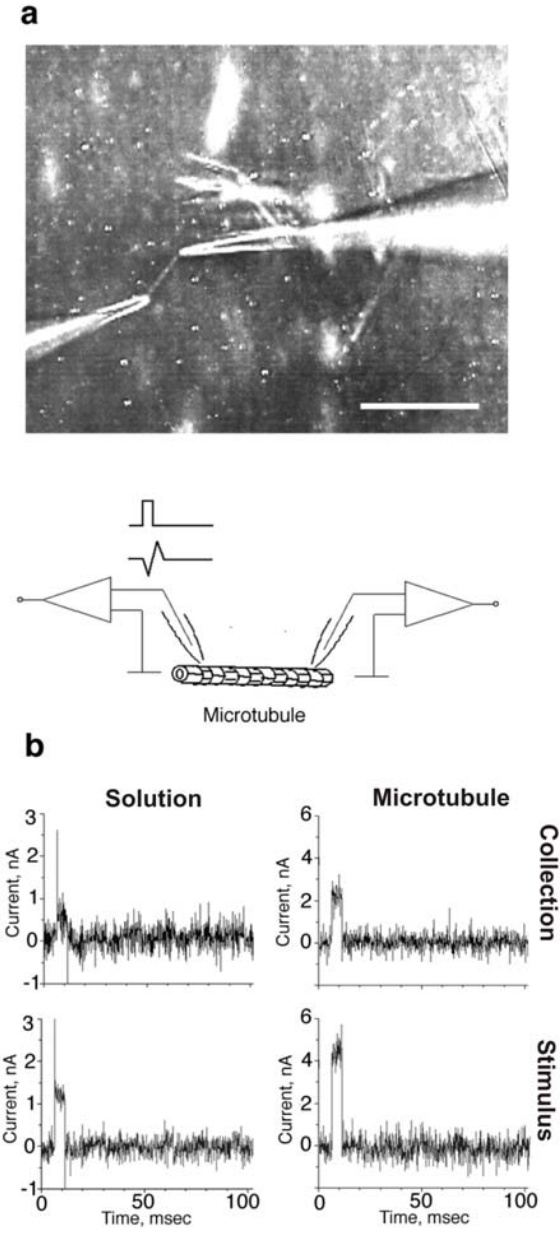

**FIGURE 1**

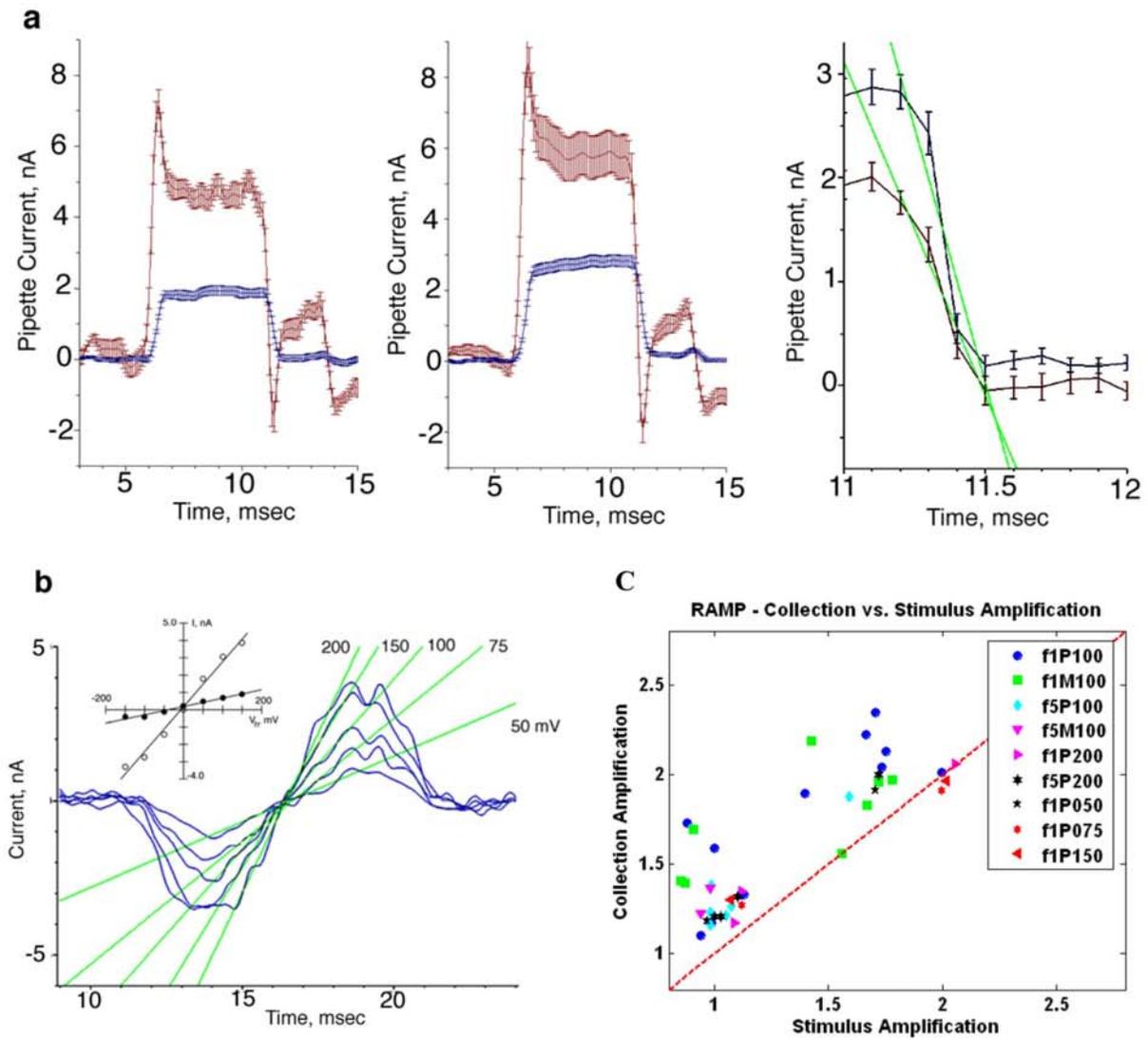

**FIGURE 2**

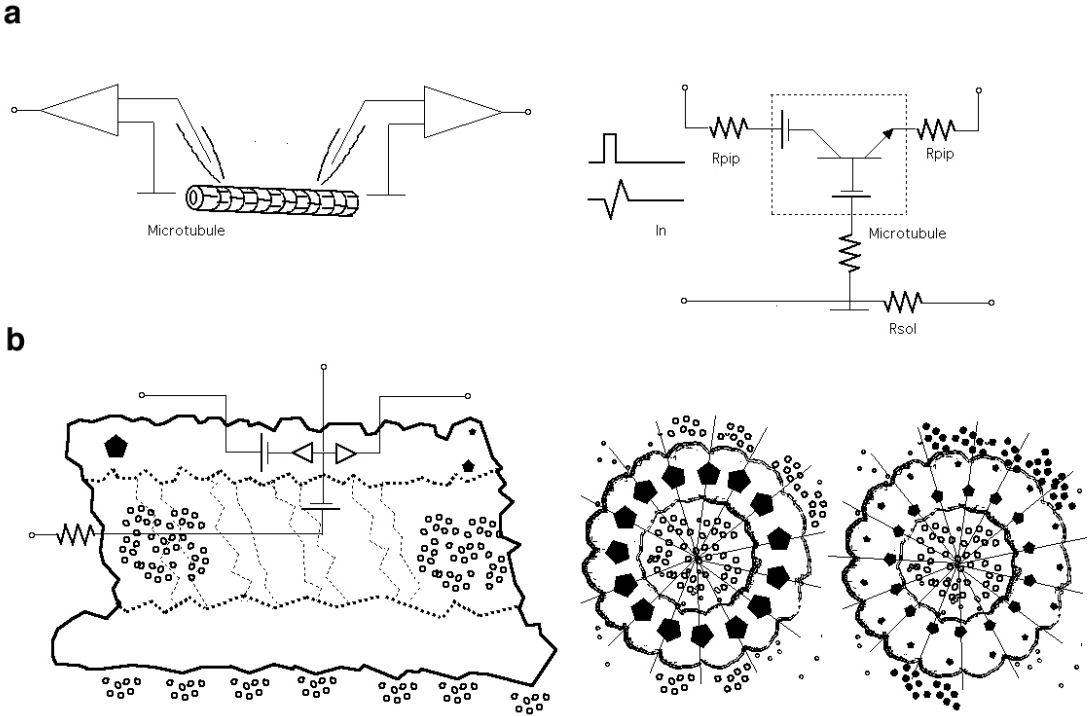

**FIGURE 3**